\documentclass{PoS}

\usepackage{amssymb, amsbsy, amsfonts}
\usepackage{amsmath, amsthm}

\let\set\mathbb

\newcommand{\MyBox}[1]{\noindent\fbox{\begin{minipage}[t]{14.8cm}
#1
\end{minipage}}}

\newcommand{\ep}{\varepsilon}

\newcounter{mmacnt}
\def\restartmma{\setcounter{mmacnt}{0}}
\restartmma \catcode`|=\active
\def|#1|{\mathrm{#1}}
\catcode`|=12
\newenvironment{mma}{
 \par\smallskip
 \catcode`|=\active
 \parskip=0pt\parindent=0pt 
 \small
 \def\In##1\\{%
   \def\linebreak{\hfill\break\null\qquad}%
   \refstepcounter{mmacnt}
   \hangindent=2.5em\hangafter=0
   \leavevmode
   \llap{\tiny\sffamily In[\arabic{mmacnt}]:=\kern.5em}%
   \mathversion{bold}\footnotesize$\tt\bf\displaystyle##1$\normalsize
   \mathversion{normal}\par
 }%
 \def\Print##1\\{%
   \def\linebreak{\hfill\break}%
   \hangindent=2.5em\hangafter=0
   \leavevmode\footnotesize ##1\par}%
 \def\Out##1\\{%
   \def\linebreak{$\hfill\break\null\hfill$}%
   \kern\abovedisplayskip\par
   \hangindent=2.5em\hangafter=0
   \leavevmode
   \llap{\tiny\sffamily Out[\arabic{mmacnt}]=\kern.5em}
   \footnotesize$\displaystyle\tt##1$\normalsize\hfill\null\par
   \kern\belowdisplayskip
 }%
 \def\Warning##1##2\\{%
   \def\linebreak{\hfill\break}%
   \hangindent=2.5em\hangafter=0
   \leavevmode
   {\scriptsize##1 : ##2}\par}%
}{%
 \par\smallskip
}

\title{\small{DESY 10-247, DO-TH 12/06, SFB/CPP-12-13, LPN 12-038   
\hfill 
}
\\ \LARGE
Evaluation of Multi-Sums for Large Scale Problems}

\ShortTitle{Evaluation of Multi-Sums for Large Scale Problems}

\author{J.~Bl\"umlein$^a$, A.~Hasselhuhn$^a$, \speaker{C.~Schneider}$^b$\\
\llap{$^a$}DESY, Zeuthen, Platanenalle 6, D-15735 Zeuthen, Germany.\\
\llap{$^b$}RISC, Johannes Kepler Universit\"at Linz, Altenberger Str. 69, A-4040 Linz, Austria.}

\abstract{
A big class of Feynman integrals, in particular, the coefficients of 
their Laurent series expansion w.r.t.\ the dimension parameter $\ep$ 
can be transformed to multi-sums over hypergeometric terms and harmonic 
sums. In this article, we present a general summation method based on 
difference fields that simplifies these multi--sums by transforming 
them from inside to outside to representations in terms of indefinite 
nested sums and products. In particular, we present techniques that 
assist in the task to simplify huge expressions of such multi-sums in 
a completely automatic fashion. The ideas are illustrated on new 
calculations coming from 3-loop topologies of gluonic massive operator 
matrix elements containing two fermion lines, which contribute to the
transition matrix elements in the variable flavor scheme.}

\FullConference{ 10th International Symposium on Radiative Corrections (Applications of Quantum Field Theory to Phenomenology) - Radcor2011\\
September 26-30, 2011\\
Mamallapuram, India}

\newtheorem{example}{Example}

\begin{document}

\section{Introduction}

\noindent
Two-point Feynman parameter integrals with one mass containing local operator insertions in $D$-dimen\-sional 
Minkowski space with one time- and $(D-1)$ Euclidean space dimensions ($\varepsilon = D - 4$ and 
$\varepsilon \in {\mathbb R}$ with $|\varepsilon| \ll 1$) can be transformed symbolically to multi--sums 
over hypergeometric terms depending on $\varepsilon$ and a discrete Mellin parameter $n$. A detailed description of the underlying algorithms was given in ~\cite{BKSF:12}.  Sums of this kind
emerge in the calulation of Feynman diagrams mentioned at 2- and 3-loops 
\cite{Blumlein:2006mh,Bierenbaum:2007qe,Bierenbaum:2007zz,BBKS:08,Bierenbaum:2009zt,HYP2,ABKSW:12} which are 
of importance for the precision measurement of the strong coupling constant and the parton distribuitions 
from the world deep-inelastic data, cf. e.g.~\cite{Alekhin:2012ig}; at the 3-loop level
a larger number of moments of these quantities has been calculated in \cite{Bierenbaum:2009mv,Blumlein:2009rg}
\footnote{However the amount of moments is far too low to reconstruct the general expressions 
depending on $n$ using the method outlined in \cite{Blumlein:2009tj}.}. 

Given such a sum 
representation like, e.g. 
in~\cite{HYP1,HYP2,BBKS:08,SchneiderSummation:10,ABKSW:12,ABHKSW:12}, the challenging task is to simplify these 
multi--sums (usually several thousands coming from Feynman integrals that describe a physical problem), in 
terms of well known special functions that can be processed further. In order to accomplish this task, a 
difference field theory for symbolic summation~\cite{Schneider:05a,Schneider:05b,Schneider:08c,Schneider:10b,Schneider:10c} 
based on $\Pi\Sigma$--fields~\cite{Karr:81} is heavily used within the summation package 
\texttt{Sigma}~\cite{Schneider:07} that generalizes the summation paradigms of~\cite{AequalB} to multi-sums. In 
addition, symbolic and analytic techniques for harmonic sums~\cite{Vermaseren:99, Bluemlein:99, 
Moch:02,ANCONT3} 
and generalizations to cyclotomic sums~\cite{ABC:11} implemented within the \texttt{HarmonicSums} package~\cite{Ablinger:12} are exploited.

In this article we report on new summation technologies that enables us to simplify huge multi--sum expressions coming from Feynman integrals to expressions in terms of indefinite nested product--sums. This allows us to represent these expressions, if possible, in terms of harmonic sums and if this is not possible,  in terms of $S$-sums or more generally in terms of cyclotomic sums; in any case, the result is given in terms of products and sums which are algebraically independent among each other~\cite{ANCONT3,Schneider:10c,ABC:11}.

The resulting ideas are implemented in the two Mathematica packages \texttt{EvaluateMultiSums} and 
\texttt{SumProduction}. The procedures are illustrated by a new calculation of fermionic contributions 
to the gluonic massive operator matrix elements ~\cite{BHS:12}.
\section{Evaluation of Multi-Sums}\label{Sec:MultiSum}

\noindent
Given a Feynman integral $F(n)$ depending on a discrete Mellin parameter $n$ and a small dimensional parameter $\ep$, one is interested in the first $s$ coefficients in its Laurent series expansion 
$$F(n)=F_{u}(n)\varepsilon^{u}+F_{u+1}(n)\varepsilon^{u+1}+\dots+F_{u+s-1}(n)\varepsilon^{u+s-1}+\dots;$$
$u\in\set Z$. As worked out in~\cite{BKSF:12}, a large class of Feynman integrals can be transformed to a 
multi--sum written in the form
\begin{equation}\label{Equ:MultiSum}
F(n)=\sum_{k_1=l_1}^{L_1(n)} ... \sum_{k_v=l_v}^{L_v(n,k_1, ..., k_{v-1})}
\sum_{k=1}^l f(\varepsilon,n,k_1,\dots,k_v)
\end{equation}
where $L_i(n,k_1,\dots,k_{v-1})$ stands for an integer linear relation in the variables $n,k_1,\dots,k_{v-1}$ or is $\infty$ and $f(\varepsilon,n,k_1,\dots,k_v)$ is an expression in terms of $\Gamma$-functions with arguments in terms of integer linear relation in the parameters $n,k_1,\dots,k_{v-1}$ and $\varepsilon$ might occur linearly in the form $r\,\varepsilon$ with $r$ being a rational number.

Then given such a sum, we will present tools to derive the required coefficients in terms of indefinite nested product-sum expressions. In particular, given this format, the coefficients can be expressed, whenever possible, in terms of harmonic sums, $S$-sums and more generally in terms of cyclotomic sums~\cite{ABC:11} defined as follows: 
\begin{equation}\label{Equ:CyclSum}
S_{(a_1,b_1,c_1),\dots,(a_s,b_s,c_s)}(x_1,\dots,x_s;n)=\sum_{i_1=1}^n\frac{x^{i_1}}{(a_1i_1+b_1)^{c_1}}\sum_{i_2=1}^{i_1}\frac{x^{i_2}}{(a_2i_2+b_2)^{c_1}}\dots\sum_{i_s=1}^{i_{s-1}}\frac{x^{i_s}}{(a_si_s+b_s)^{c_s}}
\end{equation}
with $a_i,b_i,c_i\in\set N$ ($a_i>b_i$, $c_i\neq0$) and $x_i$ being an element from the underlying field $\set K$ (containing, e.g., the complex numbers); note that choosing $a_i=1$, $b_i=0$ and $x_i\in\set\{1,-1\}$ restricts to the class of harmonic sums written in the format
$S_{x_1c_1,\dots,x_sc_s}(n)=\sum_{i_1=1}^n\frac{x_1^{i_1}}{i_1^{c_1}}\sum_{i_2=1}^{i_1}\frac{x_1^{i_1}}{i_1^{c_1}}\dots\sum_{i_s=1}^{i_{s-1}}\frac{x_s^{i_s}}{i_s^{c_s}}.$
For a formal but lengthy definition of the more general class of indefinite nested product--sums w.r.t.\ $n$ we refer to~\cite{Schneider:10b}. 
This class includes indefinite nested sums as in~\eqref{Equ:CyclSum} where also hypergeometric terms (like binomials, factorials/$\Gamma$-functions, Pochhammer symbols) might occur as polynomial expression in the numerators and denominators of the nested summands.
 
\begin{example}\label{Exp:ExpandSum}
Consider, e.g., one of the simple sums arising within the problem of~\cite{BHS:12}:
\begin{align}
F(n)&=\sum_{j_1=0}^{n-5}\sum_{j_2=0}^{n-j_1-6}\underbrace{\tfrac{\pi  2^{\ep+3} e^{-\frac{3 \gamma  \ep}{2}} (-1)^{j_1} (j_2+1) \Gamma (2-\ep) \Gamma \left(\frac{\ep}{2}+2\right) \Gamma \left(-\frac{3 \ep}{2}\right) \Gamma
   \left(-\frac{\ep}{2}+j_1+4\right) \Gamma (-j_1+n-2) \Gamma (\ep-j_1-j_2+n-5)}{(\ep-10) (\ep-8) (\ep-2) \ep \Gamma
   \left(\frac{5}{2}-\ep\right) \Gamma \left(\frac{\ep+5}{2}\right) \Gamma \left(\frac{\ep}{2}+n+1\right) \Gamma (-j_1-j_2+n-4)}}_{={f}(n,j_1,j_2)}\label{Exp:SumEp}\\[-0.8cm]
&\stackrel{?}{=}F_{-3}\varepsilon^{-3}+F_{-2}\varepsilon^{-2}+F_{-1}\varepsilon^{-1}+F_{0}\varepsilon^{0}+\dots;\nonumber
\end{align}
note that~\eqref{Exp:SumEp} is not indefinite nested w.r.t.\ $n$, since the inner sum (at least in this representation) cannot be written in the form $\sum_{j_2=l_2}^{j_1}f_2(j_2)$ where $l_2\in\set N$ and $f_2(j_2)$ is free of $j_1$ and $n$ and since in addition the outermost sum cannot be written in the form $\sum_{j_1=l_1}^{n}f_1(j_1)$ where $l_1\in\set N$ and $f_1(j_1)$ is free of $n$.
Then loading the Mathematica packages
\begin{mma}
\In << Sigma.m \vspace*{-0.06cm}\\
\Print Sigma - A summation package by Carsten Schneider
\copyright\ RISC-Linz\\
\In << EvaluateMultiSums.m \vspace*{-0.06cm}\\
\Print EvaluateMultiSums by Carsten Schneider -- \copyright\ RISC-Linz\\
\end{mma}
\noindent (and defining $f$ as the summand of our sum) we can compute the coefficients $\{F_{-3},F_{-2},F_{-1}\}$ with

\begin{mma}
\In EvaluateMultiSum[f,\{\{j_2,0,n-j_1-6\},\{j_1,0,N-5\}\},\{n\},\{5\},ExpandIn\to\{\ep,-3,-1\}]\vspace*{-0.2cm}\\
\Out \big\{0,\frac{16 (-1)^n \big(3 n^2+12 n+11\big)}{135 (n+1) (n+2)^2 (n+3)^2}
-\frac{16 \big(n^8+6 n^7-6 n^6-80 n^5-81 n^4+178 n^3+274 n^2-4 n-96\big)}{45 (n-2) (n-1)^2 n^2 (n+1) (n+2)^2 (n+3)^2}\newline
\frac{16 \big(n^2-n-8\big)}{45 (n-1) n (n+2) (n+3)}\sum_{i}^n\frac{1}{i},-\frac{8 \big(n^2-n-8\big)}{45 (n-1) n (n+2) (n+3)}\sum_{i}^n\frac{1}{i^2}+\frac{2 (-1)^n (187 n+127) \big(3 n^2+12 n+11\big)c}{2025 (n+1)^2 (n+2)^2 (n+3)^2}\newline
+\big(\frac{2 \big(17 n^6-231 n^5+121 n^4+2063 n^3-1458 n^2-2432 n+960\big)}{675 (n-2) (n-1)^2 n^2 (n+1) (n+2) (n+3)}-\frac{16 (-1)^n \big(3 n^2+12 n+11\big)}{135 (n+1) (n+2)^2 (n+3)^2}\big) \sum_{i}^n\frac{1}{i}\newline
+\tfrac{2 \big(43 n^{12}+112 n^{11}+263 n^{10}-216 n^9-11309 n^8-16476 n^7+55837 n^6+78164 n^5-95178 n^4-116688 n^3+51784 n^2+30624 n-23040\big)}{675 (n-2)^2 (n-1)^3 n^3 (n+1)^2 (n+2)^2 (n+3)^2}\big\}.\\
\end{mma}
\noindent The input $\{n\}, \{5\}$ means that we suppose that $n\geq5$; this specification of the parameter range is necessary for the internal calculations.
Loading in addition J.~Ablinger's \texttt{HarmonicSums} package~\cite{Ablinger:12} would automatically transform the output sums to the harmonic sums $S_1(n),S_2(n)$. The constant term $F_0(n)$ (which would be too large to print here) introduces in addition the sum $S_3(n)$ and the zeta-value $\zeta_2$ defined by the Riemann zeta function $\zeta_z=\sum_{i=1}^{\infty}z^{-i}$; the total timings (including the constant term) are 4 minutes.
\end{example}

To get these coefficients, we expand the summand of~\eqref{Equ:MultiSum} in the parameter $\varepsilon$
$$f(\varepsilon,n,k_1,\dots,k_v)=f_{u}(\varepsilon,n,k_1,\dots,k_v)\varepsilon^{u}+f_{u+1}(\varepsilon,n,k_1,\dots,k_v)\varepsilon^{u+1}+\dots$$
by using formulas such as
$\Gamma(n+1+\bar{\ep}) = \frac{\Gamma(n) \Gamma(1+\bar{\ep})}{B(n,1+\bar{\ep})}$
with $\bar{\ep} = r \ep$ for some $r\in\set Q$ and

\vspace*{-0.2cm}

\begin{equation*}
B(n, 1 + \bar{\ep}) = \frac{1}{n}\exp\left(\sum_{k=1}^\infty \frac{(-\bar{\ep})^k}{k} S_k(n)\right)
= \frac{1}{n}\sum_{k=0}^\infty (-\bar{\ep})^k S_{\underbrace{\mbox{\scriptsize 1, \ldots
,1}}_{\mbox{\scriptsize
$k$}}}(n)
\end{equation*}

\vspace*{-0.2cm}

\noindent and other well-known transformations for the $\Gamma$-functions. 
Finally, under the assumption that the sums are uniform convergent, in particular if one can exchange summation signs and differentiation, one gets
the coefficients

\vspace*{-0.4cm}

$$F_i(n)=\sum_{k_1=l_1}^{L_1(n)} ... \sum_{k_v=l_v}^{L_v(n,k_1, ..., k_{v-1})}
f_{i}(\varepsilon,n,k_1,\dots,k_v)$$

\vspace*{-0.1cm}

\noindent of the expansion~\eqref{Equ:MultiSum}.
Now the hard task is to simplify these sums further.
Here we rely on the following summation paradigms~\cite{Schneider:07} in the context of difference fields.

\medskip

\MyBox{\noindent\textbf{Deriving recurrences}:
\noindent{\it
Given} an integer $d\geq0$ and given a sum

\vspace*{-0.3cm}

\begin{equation}\label{Equ:CreaSum}
F(\vec{m}, n):=\sum_{k=l}^{L(\vec{m},n)}f(\vec{m},n,k)
\end{equation}

\vspace*{-0.2cm}

where $l\in\set N$, $f(\vec{m},n,k)$ is an expression in terms of indefinite nested product--sums w.r.t.\ $k$, and where $\vec{m}=(m_1,\dots,m_r)$ and $n$ are discrete parameters; {\it find} 

\vspace*{-0.3cm}

\begin{equation}\label{Equ:rec}
a_0(\vec{m},n)F(\vec{m},n)+\dots+a_d(\vec{m},n)F(\vec{m},n+d)=h(\vec{m},n).
\end{equation}

\vspace*{-0.1cm}

where $a_0(\vec{m},n),\dots,a_d(\vec{m},n)$ are rational functions in $\vec{m}$ and $n$, and $h(\vec{m},n)$ consists of an expression in terms of sums of the type as $F(\vec{m},n)$, but with simpler summands (i.e., less summation objects or less nested summation quantifiers).}

\medskip

\noindent \text{Remarks.} \textbf{(1)}~The underlying algorithms~\cite{Schneider:08c} utilize creative telescoping that has been originally introduced with Zeilberger's algorithm~\cite{Zeilberger:91} for hypergeometric summands.\\
\textbf{(2)} The upper bound $L(\vec{m},n)$ might depend integer linearly on $n,\vec{m}$ or might be $\infty$. In this particular case, in order to get~\eqref{Equ:rec}, limit calculations are necessary using asymptotic expansions of the summand objects. For the expansion of harmonic sums, $S$-sums and cyclotomic sums efficient algorithms are developed~\cite{ABC:11} and implemented in \texttt{HarmonicSums}~\cite{Ablinger:12}.

\medskip

\noindent Now suppose that we are given a recurrence of the type~\eqref{Equ:rec} where we succeed in representing $h(\vec{m},n)$ in terms of indefinite nested product--sums w.r.t.\ $n$. Then we can proceed as follows.

\smallskip

\MyBox{\noindent\textbf{Recurrence solving}:
\noindent{\it Given} such a simplified 
recurrence, {\it find} all solutions that are expressible in terms of indefinite nested product--sum expressions w.r.t.\ $n$.}

\medskip

\noindent\textit{Remark.} The underlying algorithms, see e.g.~\cite{Abramov:94,Schneider:05a,ABPS:12}, find indefinite nested sum expressions that are highly nested: the maximal nesting depth will be the recurrence order. Thus the simplification to expression with minimal  depth is a crucial step~\cite{Schneider:10b}. In addition, the occurring sums are reduced, i.e., they are algebraically independent among each other~\cite{Schneider:10c}.

\medskip

\MyBox{\noindent\textbf{Combination of solutions:} \textit{Given} these simplified solutions, \textit{combine} them such that the evaluation agrees with the original sum $F(\vec{m},n)$ for the first, say $n=l,...,l+d-1$ initial values.}

\medskip

\noindent If this is possible, it follows --up to some mild side conditions-- that this combination of indefinite nested product--sums is equal\footnote{We emphasize that in each computation step of this transformation we produce proof certificates which enables one to verify rigorously the correctness of the result.} to the definite input sum for all $n\geq l$.

\begin{example}
The goal is to find a representation in terms of indefinite nested product-sums w.r.t.\ $n$ for the coefficient 
\vspace*{-1.2cm}

\begin{equation}\label{Equ:FM2Coeff}
F_{-2}(n)=\sum_{j_1=0}^{n-5}\tfrac{-8(j_1+2) (j_1+3) (n+1) (j_1-n+3) (j_1-n+4)(-1)^{j_1} (j_1+1)! (-j_1+n-5)!}{135(n+1)!}\overbrace{\sum_{j_2=0}^{n-j_1-6}\frac{j_2+1}{j_1+j_2-n+5}}^{=f(n,j_1)}
\end{equation}

\vspace*{-0.2cm}

\noindent in the expansion of sum~\eqref{Exp:SumEp}. In the first round,  we transform the summand $f(n,j_1)$ to indefinite nested product--sums w.r.t.\ $j_1$ with the algorithms implemented in \texttt{Sigma}. First, we compute a recurrence for $f(n,j_1)$:

\vspace*{-0.4cm}

$$(-j_1+n-5)f(n,j_1)+(j_1-n+4)f(n,j_1+1)=j_1-n+5.$$

\vspace*{-0.06cm}

\noindent Next, we solve the recurrence and get the full solution space
$L=\{ c(-j_1+n-4)+
(-j_1+n-4) \sum_{i=1}^{j_1} \frac{1}{-3+n-i}|c\in\set R\}$
which means that $f(n,j_1)\in L$. 
Now we have to combine the solutions (i.e., to determine $c$) so that they agree with $f(n,j_1)$ at, e.g., $j_1=0$, i.e., $f(n,0)=\sum_{j_2=0}^{n-j_1-6}\frac{j_2+1}{j_2-n+5}$. Again, we turn this sum to an indefinite nested sum (by computing a recurrence, solving the recurrence and combining the solutions) and obtain 
$f(n,0)=\frac{(n-4)\big(n^4-2 n^3-7 n^2+16 n-6\big)}{(n-3) (n-2) (n-1) n}+(4-n)\sum_{i=1}^n\frac{1}{i}$. Choosing $c=f(n,0)/(n-4)$ finally yields

\vspace*{-0.7cm}

$$f(n,j_1)=(-j_1+n-4) \sum_{i=1}^{j_1} \frac{1}{-3+n-i}+\tfrac{\big(n^4-2 n^3-7 n^2+16 n-6\big) (-j_1+n-4)}{(n-3) (n-2) (n-1) n}+(j_1-n+4)\sum_{i=1}^n\frac{1}{i}.$$

\vspace*{-0.2cm}

\noindent With this indefinite nested sum representation w.r.t.\ $j_1$ of the summand $f(n,j_1)$, we start the final round: we compute a recurrence for~\eqref{Equ:FM2Coeff}, solve the recurrence and combine the solutions to get the coefficient $F_{-2}(n)$  computed in Example~\ref{Exp:ExpandSum}.
\end{example}

In a nutshell, given a multi--sum of the form

\vspace*{-0.3cm}

$$F(\vec{m},n)=\sum_{k=l}^{L(\vec{m},n)}\underbrace{\sum_{k_1=l_1}^{L(\vec{m},n,k)} ... \sum_{k_v=l_v}^{L_v(\vec{m},n,k,k_1, ..., k_{v-1})}
\overline{f}(\vec{m},n,k,k_1,\dots,k_v)}_{f(\vec{m},n,k)}$$

\vspace*{-0.2cm}

\noindent where $\overline{f}$ itself is an expression in terms of indefinite nested products-sums (in particular the class of sums given in~\eqref{Equ:MultiSum} is covered), we apply the following method to transform $F(\vec{m},n)$ 
to an expression in terms of indefinite nested product--sums.

\vspace*{-0.2cm}

\begin{enumerate}
\item Transform the outermost summand $f(\vec{m},n,k)$ to an expression in terms of indefinite nested product--sums w.r.t.\ $k$ by applying the method recursively to all the arisings definite sums (i.e.,  the parameter vector $\vec{m}$ is replaced by $(n,\vec{m})$ and the the role of $n$ is $k$). Note that the occurring sums in $f$ are simpler than $F(n)$ (one definite sum less). If the summand is free of sums, nothing has to be done.

\item Compute a recurrence~\eqref{Equ:rec} for the sum~\eqref{Equ:CreaSum};
note that by construction the right hand side might be again an expression in terms of definite multi--sums, but these summands are of simpler format than $f$. Apply the method recursively to these sums and compute a right hand side representation which consists of indefinite nested product--sums w.r.t.\ $n$.

\item Solve the recurrence~\eqref{Equ:rec} in terms of indefinite nested product--sums w.r.t.\ $n$.

\item Compute $d$ initial values (i.e., specialize the parameter $n$ to appropriate values from $\set N$, say $n=l,l+1,\dots,l+d-1$, and apply the method recursively to the arising sums where $m_1$ takes over the role of $n$ and the parameters are $\vec{m}$. If no parameter is left, the expression is a constant. In particular, if there is no sum left, nothing has to be done. Otherwise, in the outermost summations the upper bound is $\infty$. In this case, our method is applied to transform the summands to indefinite nested product--sums w.r.t.\ the outermost summation index. Usually, the derived sums can be transformed to multiple zeta values~\cite{MZV} or infinite versions of $S$--sums or cyclotomic sums~\cite{ABC:11}. Otherwise, they are kept in the given format.

\item Try to combine the solutions to find an indefinite nested product-sum representation w.r.t.\ $n$ of $F(\vec{m},n)$. If this fails, ABORT.
\end{enumerate}

\noindent\textit{Remarks.} (1) The \textit{existence} of a recurrence in step~2 is guaranteed by using arguments form~\cite{AequalB} and~\cite{Mallinger:96}. Only
computation issues are a bottleneck. Usually, we succeed in finding recurrences when $f$ consists up to 100 indefinite nested product--sum objects. If $f$ is more complicated (or if it seems appropriate), the sum is split into several parts and the method is applied separately.\\
(2) \textit{Termination:} The method is applied recursively to sums which are always simpler than the original sum (less summation quantifiers, less parameters,  or less objects in the summand). After finitely many recursion steps, one arrives at the base cases where no summation quantifiers arise.\\
(3) \textit{Success:} If the method does not abort in step~5, it terminates and outputs an indefinite nested product-sum expression w.r.t.\ $n$. As a consequence, finding not sufficiently many solutions of a given recurrence of the type~\eqref{Equ:rec} in step~5 is the only reason why the method might fail. For general multi--sums this failure would happen all over. However in the context of Feynman integrals, we found almost always $d$ linearly independent solutions of the homogeneous version~\eqref{Equ:rec} and one particular solution of the recurrence itself; in these cases, the solution space of~\eqref{Equ:rec} is completely determined, and the failure in step~5 cannot occur. The (for us surprisingly) rare case that not sufficiently many solutions are found was always an indication that the sum representation of the Feynman sums could be improved so that afterwards our method worked.\\
(4) \textit{Subtle details:} The input sums coming from Feynman integrals are rather challenging. In each step of the method, pole issues arose (lower  and upper bounds must be updated during the calculations and thus compensating terms must be computed separately, the initial values must be chosen carefully, etc.). As a consequence, the 5 line method above implemented in the package \texttt{EvaluateMultiSums} requires currently about 8000 lines of code in Mathematica (not counting the implementations for recurrence finding and solving which are part of the package \texttt{Sigma}).

The package \texttt{EvaluateMultiSums} has been applied successfully to various complicated 3--loop ladder 
graphs~\cite{HYP1,SchneiderSummation:10}. In particular, the graphs from~\cite{ABHKSW:12} could be calculated 
by reducing it to several multi--sums, one of them being
\begin{align*}
\sum_{j=0}^{n-3}& \sum_{k=0}^j \sum_{l=0}^k \sum_{q=0}^{-j+n-3} \sum_{s=1}^{-l+n-q-3} \sum_{r=0}^{-l+n-q-s-3}\tfrac{\binom{j+1}{k+1}
\binom{k}{l} \binom{n-1}{j+2} \binom{-j+n-3}{q} \binom{-l+n-q-3}{s}
\binom{-l+n-q-s-3}{r} r! (-l+n-q-r-s-3)! (s-1)!}{(-l+n-q-2)!(-j+n-1)(n-q-r-s-2) (q+s+1)}\\[-0.1cm]
&(-1)^{-j+k-l+n-q-3}\Big[4 S_1(-j+n-1)-4S_1(-j+n-2)-2S_1(k)-(S_1(-l+n-q-2)\\[-0.1cm]
&\hspace*{3cm}+S_1(-l+n-q-r-s-3)-2 S_1(r+s))+2 S_1(s-1)-2S_1(r+s)\Big].
\end{align*}
\section{Mass production}

\noindent
For various problems, like e.g.~\cite{HYP1,HYP2}, several thousand sums have to be evaluated. In particular, for the brand new calculation of fermionic contributions 
to the gluonic massive operator matrix elements ~\cite{BHS:12} the $\ep$-expansion of a 2 GByte expression  
consisting of 2419 multi--sums was calculated; one of the simple sums is, e.g.,~\eqref{Exp:SumEp}. With a 
lot of computer resources this problem could be tackled~\footnote{For testing we calculated around 800 of these 
sums with the help of 16 Mathematica processes in around 4 days.}. However, we can do it much better by using the following routines which are available in the new package 

\begin{mma}
\In << SumProduction.m \\
\Print SumProduction - A summation package by Carsten Schneider \copyright\ RISC-Linz\\
\end{mma}

\medskip

\noindent \textbf{1. Reduction to key sums:} First, we reduce the 2 GByte expression (stored in \texttt{expr}) to key sums with the function call
\begin{mma}
\In compactExpr=ReduceMultiSums[expr,\{n\},\{5\}];\\
\end{mma}
\noindent The reduced expression \texttt{compactExpr} is only 7.6 MByte large and contains only 29 sums and 15 terms free of sums; in total it took us 6 hours and 53 minutes to obtain this reduction.

\medskip

\noindent\textit{Remark.} Internally, the 2419 sums have been synchronized w.r.t.\ the occurring summation ranges (taking for each class the maximum of the lower bounds and the minimum of the upper bounds). As result, we obtained only 4 sums with equalized summation ranges
$$\sum_{i_2=5}^{n-5}\sum_{i_1=0}^{i_2}h_1(\varepsilon,n,i_2,i_1),\;\;\;\sum_{i_2=0}^{n-5}\sum_{i_1=0}^{n-i_2-5}h_2(\varepsilon,n,i_2,i_1),\;\;\;\sum_{i_1=5}^{n-5}h_3(\varepsilon,n,i_1),\;\;\;\sum_{i_1=0}^{\infty}h_4(\varepsilon,n,i_1)$$
plus a large term free of summation quantifiers.
Next, all the occuring Pochhammer symbols, factorials/$\Gamma$-functions, and binomials are written in a basis of algebraically independent objects plus the extra object $(-1)^n$ (if necessary); for details see~\cite[Sec.~9]{Schneider:10c} and~\cite[Sec.~6]{Schneider:05b}. 
Finally, the expressions are split further to get the form $\sum h(n,(i_2,)i_1,\ep)*r(n,(i_2,)i_1,\ep)$ or $h(n,\ep)*r(n,\ep)$ where $h$ stands for a (proper) hypergeometric term in $n$ (and $i_1,i_2$), i.e., being a product of binomials/factorials/Pochhammers in the numerator and denominator, and $r(n,(i_2,i_1),\ep)$ being a rational function in $n,\ep$ (and $i_1,i_2)$; note that $r$ might fill several pages.

\medskip

\noindent\textbf{2. Computing $\ep$--expansions in parallel:} Next, we compute the coefficients of the $\ep$-expansion as outlined for sum~\eqref{Exp:SumEp}; as it turns out, the time to calculate the expansion of a key sum is 
similar to calculating just one typical candidate within the 2419 sums contributing to the corresponding key sum. In order to produce these expansions automatically, we developed the routine

\begin{mma}
\In ProcessEachSum[compactExpr,\{n\},\{6\},ExpandIn\to\{ep,-3,0\}]\\
\end{mma}

\noindent which sequentially applies \texttt{EvaluateMultiSum} with the corresponding input parameters to the occurring multi-sums in \texttt{compactExpr}. This step took in total 2 hours and 35 minutes.

\medskip

\noindent\textit{Remark.}
Internally, it takes the first multi-sum and generates a file with the name SUM1. If the result is computed, the file is updated with the result. Then the routine continues with the second sum provided the file SUM2 is not existent on the hard disk. In this way, ProcessEachSum can be executed in parallel for mass productions (for even larger problems than this).

\medskip

\noindent\textbf{3. Combining the subresults:} Finally, the coefficients of the expansions of the subresults are read from the hard disk and are summed up to the final coefficients of the expansion:

\begin{mma}
\In result=CombineExpression[compactExpr,\{n\},\{6\}];\\
\end{mma}

\noindent\textit{Remark.} Internally, the expressions are reduced further by eliminating all algebraic relations of the occurring sums and products~\cite{ANCONT3,Schneider:10c,ABC:11}. This step took 21 seconds.

\medskip

The final result can be expressed in terms of 
$$\zeta_2,\zeta_3,(-1)^n,S_1(n),S_2(n),S_3(n),S_{2,1}(n),S_{3,1}(n),S_{2,1,1}(n)$$
and requires about 100 KByte memory.
The total calculation took around 9 hours and 30 minutes. 

\medskip

In conclusion, we are ready to go into mass production for various challenging problems being of 
similar type as outlined in~\cite{HYP2,ABKSW:12,ABHKSW:12,BHS:12}.

\vspace*{2mm}
\noindent
{\bf Acknowledgment.}
This work has been supported in part by DFG Sonderforschungsbereich Transregio 9, Computergest\"utzte 
Theoretische Teilchenphysik, Austrian Science  Fund (FWF) grant P203477-N18, and EU Network {\sf LHCPHENOnet} 
PITN-GA-2010-264564. 


\end{document}